\documentclass{an}
\usepackage{graphicx}
\usepackage{times}
\usepackage{fancyhdr}
\begin{document}

\title{Nanodiamond dust and the energy distribution of quasars}

\author{L. Binette\inst{1} \and A. C. Andersen\inst{2}  \and
H. Mutschke\inst{3} \and S. Haro-Corzo\inst{1}}
\institute{
Instituto de Astronom\'\i a, 
Universidad Nacional Aut\'onoma de 
M\'exico, Apartado Postal 70-264, 04510 M\'exico,  DF, Mexico
\and 
Dark Cosmology Center, Juliane Maries Vej 30, DK-2100 Copenhagen, Denmark
\and 
Astrophysikalisches Institut und Universit\"ats-Sternwarte (AIU), Schillerg\"a\ss chen 3, 
D-07745 Jena, Germany}


\abstract{The spectral energy distribution of quasars shows a sharp
steepening of the continuum shortward of $\simeq 1100$\,\AA. 
The steepening could be a result of dust absorption. 
We present a dust extinction model 
which considers crystalline carbon grains and compare it 
with SMC-like dust extinction consisting of a mixture of
silicate grains with graphite or amorphous carbon grains.
We show that the sharp break seen in {\it individual} quasar spectra of 
intermediate 
redshif $\sim 1$--2 can be reproduced by dust absorption provided the
extinction curve consists of nanodiamonds, composed of terrestial cubic 
diamonds or of diamonds similar to the 
presolar nanodiamonds found in primitive meteorites. 
}

\correspondence{anja@nordita.dk}

\maketitle

\section{Introduction}

The ultraviolet energy distribution of quasars is characterized by the
so-called ``big blue bump'', which peaks in ${\nu}F_{\nu}$ at
approximately 1000\,\AA.  The quasar `composite' spectral energy
distribution (SED) of Telfer et\,al. (2002, hereafter TZ02) obtained
by co-adding 332 HST-FOS archived spectra of 184 quasars between
redshifts 0.33 and 3.6, exhibits a steepening of the continuum at
$\sim 1100$\,\AA. A fit of this composite SED using a broken powerlaw
reveals that the powerlaw index changes from $-0.69$ in the near-UV to
$-1.76$ in the far-UV (see TZ02). We label this observed sharp
steepening the `far-UV break'.  In these proceedings, we compare the
dust absorption that results from nanodiamonds with that of more
traditional dust species. In a previous proceedings (Binette
et\,al. 2005b), we described how we came to consider the possibility
of carbon crystallite dust. The argumentation in support of the dust
absorption interpretation of the UV break has been fully described in
Binette et\,al. (2005a, hereafter BM05). Complementary information
about crystalline dust can be found in this and other proceedings
(Binette et\,al. 2005b, 2005c).

\section{Dust extinction model}\label{sec:dust}

\begin{figure}
\resizebox{\hsize}{!}
{\includegraphics[width=\columnwidth,height=8cm]{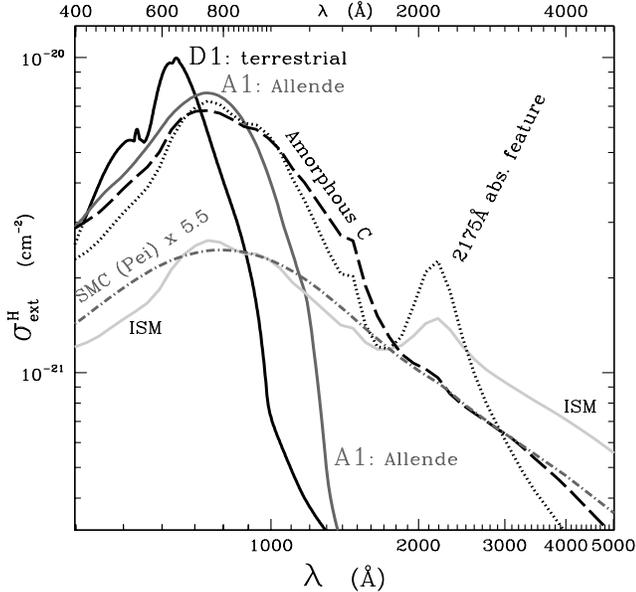}}
\caption{Extinction cross section for terrestrial cubic nanodiamonds 
(black continuous line labeled D1) and meteoritic nanodiamonds from
the Allende Meteorite (gray continuous line labeled A1).  The
silver continuous line represents a graphite and silicate model of the ISM
dust by Martin \& Rouleau (1991), with grain sizes ranging from 50 to
2500\,\AA.  The dotted line corresponds to a similar (ISM) grain
composition but with sizes confined to the range 50 to 300\,\AA.  Notice the
graphite absorption feature at 2175\,\AA\ in both currves. The long
dashed line describes a dust model consisting of grains composed of
silicate (50\%), amorphous carbon (45\%) and 5\% of graphite, with
grain sizes within the range 50 to 300\,\AA. The dot-dashed gray line
represents the SMC dust model of Pei (1992) multiplied here by the
factor 5.5. }
\label{fig:ext}
\end{figure}

Nanodiamonds are the most abundant presolar grain type found in the
relatively unprocessed meteorites, called carbonaceous chondrites. The
diamonds account for about 3\% of the total amount of carbon in this type of
meteorites! The main difference between meteoritic diamonds and cubic
terrestrial diamonds is that meteoritic diamonds possess surface
impurities that significantly alter  their optical properties. This can
be appreciated in Fig.\,\ref{fig:ext} where we compare the extinction
cross-section resulting from cubic  terrestrial diamonds (black continuous
line) with that from nanodiamonds from the Allende meteorite (gray
continuous line). In both dust models, we assume a powerlaw grain size
distribution of index $-3.5$ bounded by size limits of 2 and 25\,\AA.
We  assume the grains to be spherical  and adopt the Mie
theory (Bohren and Huffman 1983)  and the complex refraction indices $n+ik$ 
of  Mutschke et\,al. (2004) (for the Allende nanodiamonds) 
and Edwards \& Philipp (1985)  (for the terrestrial diamonds). 
It turns out that within  the above small grain size
regime, using a log-normal distribution or altering  the  above grain size
limits slightly, do not produce any differences in the derived extinction
curves, as explained in BM05.  

It is apparent from Fig.\,\ref{fig:ext} that the steepness of the {\it
near-UV} extinction rise in the case of nanodiamonds ought to produce
a sharp absorption break, a desired feature for any dust model that
aims at fitting the UV quasar break.  The nanodiamond extinction
curves in Fig.\,\ref{fig:ext} were normalized in such a way that they
represent the case of having all the carbon in the dust (assuming a
solar C/H abundance ratio). This normalization would need to be scaled
according to the actual but unknown dust-to-gas ratio appropriate to
the quasar environment.

\begin{figure}[!t]
  \includegraphics[width=\columnwidth]{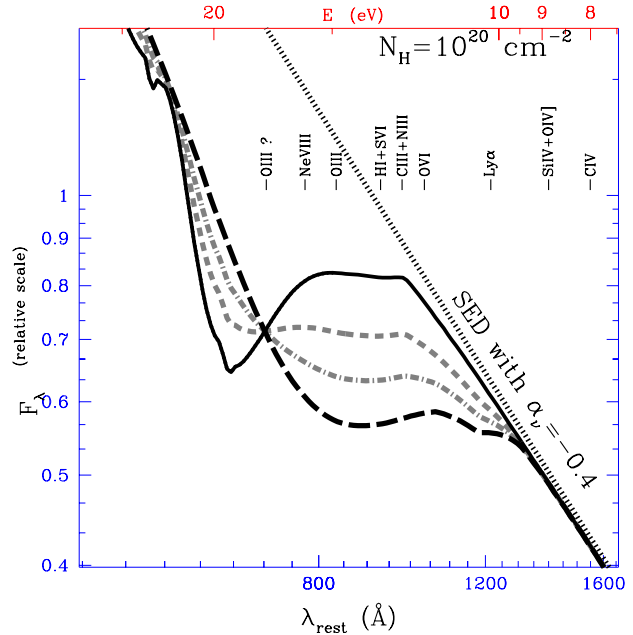}
  \caption{Absorption models assuming a powerlaw SED of index
  $\alpha_{\nu}=-0.4$ (dotted line). The nanodiamond dust screen corresponds
  to a hydrogen column of $10^{20}$\,cm$^{-2}$.  The absorbed SED
  represented by the continuous line corresponds to the extinction by terrestrial cubic
  diamonds (D1), while the long dashed line corresponds to nanodiamonds from
  the Allende meteorite (A1). The gray lines correspond to a mixture of the
  two flavors: dashed line: 60\% D1 + 40\% A1, dot-dashed line: 30\%
  D1 + 70\% A1.  Labelled pointers indicate where emission lines are
  expected in the rest-frame quasar spectra.}
\label{fig:abs}
\end{figure}

\section{Powerlaws absorbed by nanodiamond dust}\label{sec:abs}

In BM05, we present calculations in which the dust is either intrinsic
to the quasars or intergalactic. In the end, we concluded that only
the {\it intrinsic dust} hypothesis was satisfactory. The calculations
presented here will assume the intrinsic dust case, which requires gas
columns of order $10^{20}$\,cm$^{-2}$ assuming Solar carbon abundance and full
depletion onto nanodiamond grains.

Both nanodiamond types can produce a sharp UV break as shown in
Fig.\,\ref{fig:abs} where we assumed a powerlaw energy distribution of
spectral index $-0.40$ (dotted line) and a column of
$10^{20}$\,cm$^{-2}$. The opacity is given by $\tau_{\lambda}^{ext} =
N_{\rm H} \sigma^{H}_{ext}(\lambda)$ and the transmission function is
simply $e^{-\tau_{\lambda}^{ext}}$. It is interesting to notice that
the break in the absorbed powerlaws in Fig.\,\ref{fig:abs} is
shortward (longward) of Ly$\alpha$ in the case of cubic diamonds
(meteoritic nanodiamonds), respectively.

\begin{figure*}[!t]
\includegraphics[width=\columnwidth,height=8cm]{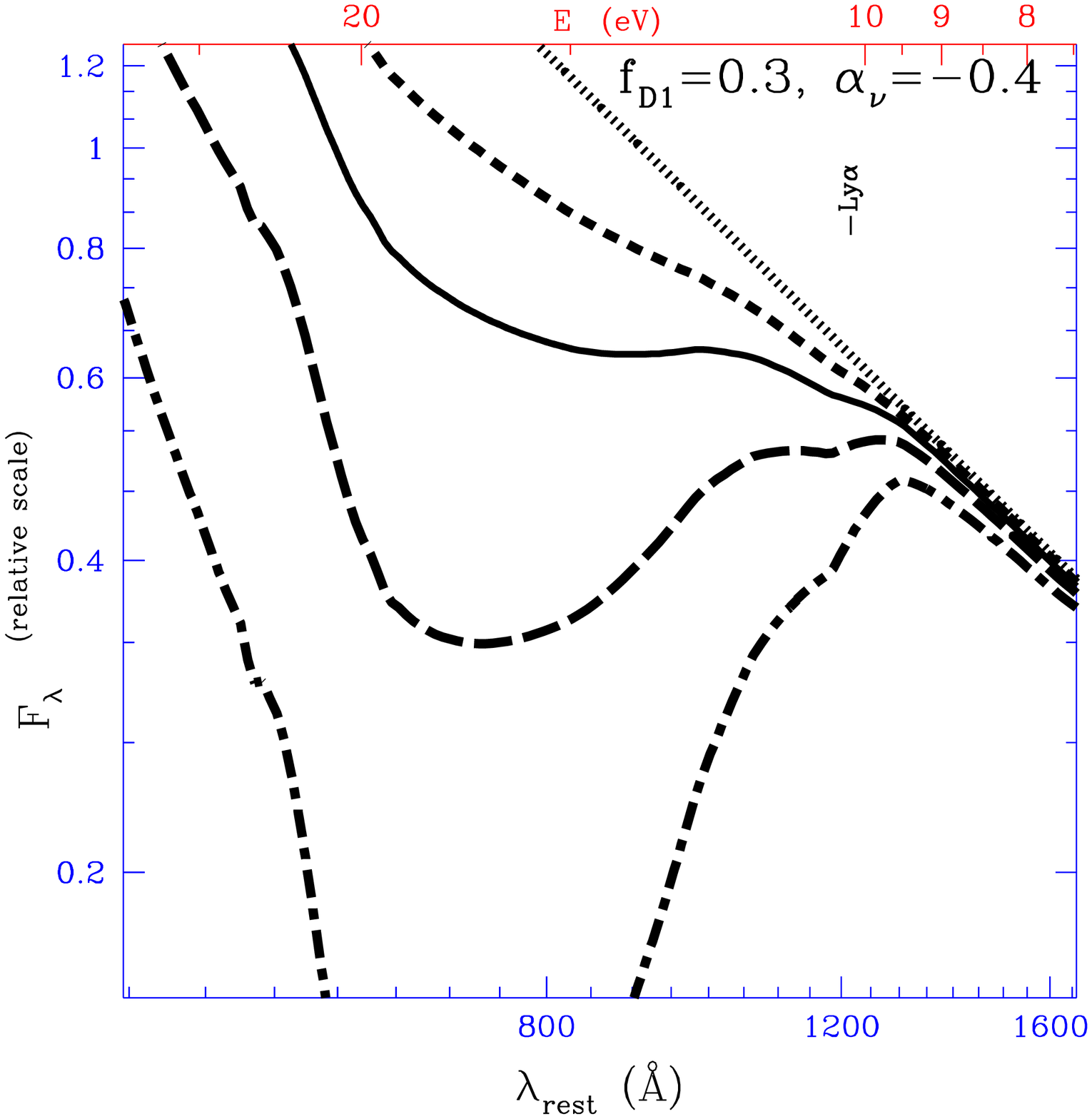}%
\hspace*{\columnsep}%
\includegraphics[width=\columnwidth,height=8cm]{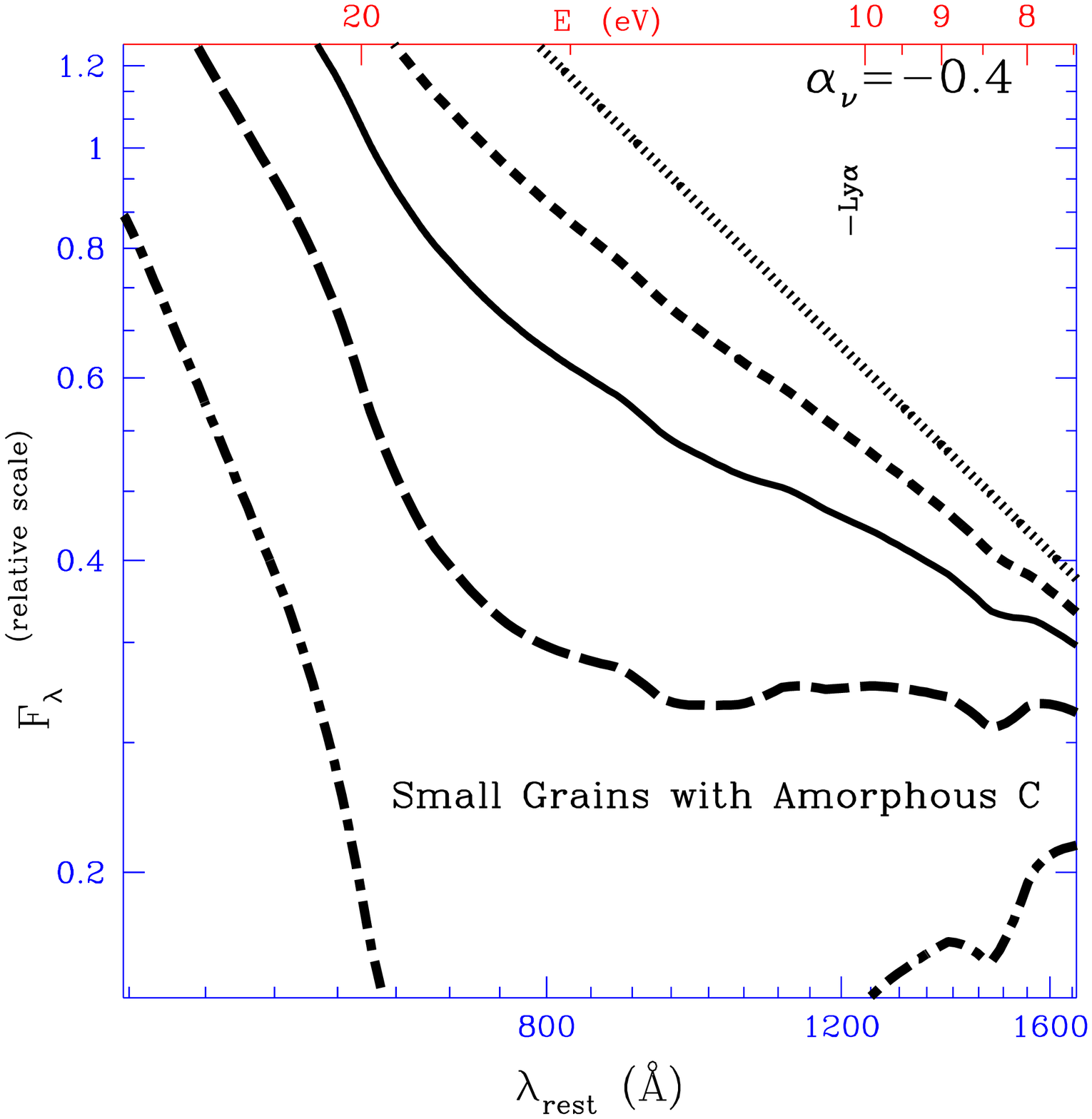}
\caption{Absorbed powerlaw SED of index $\alpha_{\nu}=-0.4$ (dotted line)
for the following $N_{\rm H}$  columns in units of $10^{20}$\,cm$^{-2}$:
$N_{20}=0.5$ (dashed line), 1.0 (solid line), 2.0 (long dashed line) and 4.0
 (dash-dot line). The following two dust compositions are
considered. Panel a (left): the dust corresponds to a combination of
terrestrial cubic diamonds (30\%) and primitive meteorite nanodiamonds
(70\%). Panel b (right): the dust grains are  made of
silicate (50\%), amorphous carbon (45\%) and graphite (5\%).  }
\label{fig:dust}
\end{figure*}

It turns out that the UV break position in most spectra is about midway between the break positions
produced by  either of the two nanodiamond types, suggesting
that a combination of both types is needed, as
described in more detail in BM05. We found that most quasar spectra favor a
composition consisting on the order of 30\% of cubic diamonds and 70\% of
meteoritic nanodiamonds.  In Fig.\,\ref{fig:dust}a, we illustrate the
effect of varying the absorption column $N_{\rm H}$, assuming the above mixture
(30\% and 70\%, respectively, equivalent to $f_{D1}=0.3$). The curves shown
correspond to the four H columns values of $N_{20}=0.5$, 1.0, 2.0 and 4.0, in
units of $10^{20}$\,cm$^{-2}$. The same powerlaw SED  (dotted line) is assumed as for 
Fig.\,\ref{fig:abs}, that is $\alpha_{\nu}=-0.4$.

\begin{figure}
\resizebox{\hsize}{!}
{\includegraphics[width=7cm,height=7cm]{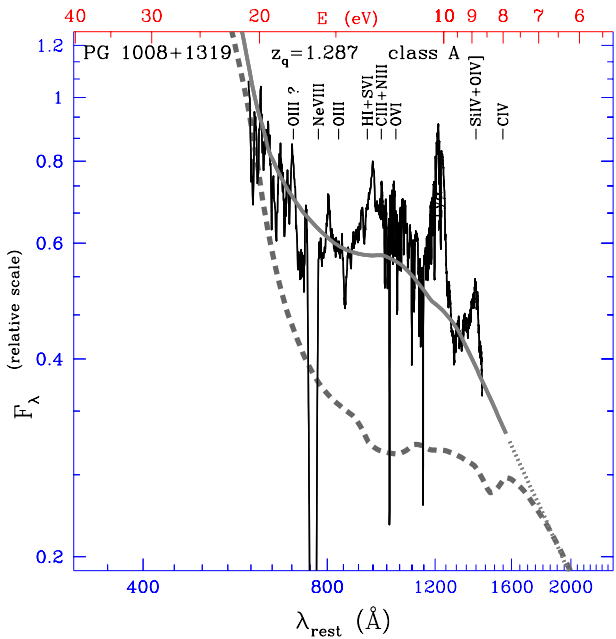}}
\caption{The rest-frame spectrum of class (A) quasar PG\,1008+1319.  Notice the far-UV rise
shortward of the UV break. The nanodiamond dust extinction  model (gray continuous
line) assumes a powerlaw with $\alpha_{\nu}=+0.13$ (from Neugebauer
et\,al. 1987), $N_{20}=1.2$ and the same dust composition as in
Fig.\,\ref{fig:dust}a (mixed nanodiamond types).  The gray dashed line corresponds to extinction by 
silicate-amorphous carbon dust  (as in Fig.\,\ref{fig:dust}b)
assuming a column $N_{20}=2.4$. The transmitted flux in the latter
model has been multiplied by 1.3 so that longward of 1800\,\AA\ it
superimposes the transmitted flux of the nanodiamond model. }
\label{fig:pg}
\end{figure}

\section{Comparison with SMC-like dust}\label{sec:smc}

Can ISM or SMC-like dust reproduce the quasar break? According to
Shang et\,al. (2005), reddening by ISM or SMC-like grains ``is not
able to produce the spectral break seen in the AGN sample, without
leaving a clear signature at longer wavelengths''. Instead of a sharp
break, ISM extinction results in a shallow rollover that extends in
the near-UV and even in the optical domain. In the HST-FOS sample that
we studied, most spectra (the 50 class A and B spectra) are quite hard
in the near-UV, consistent with a powerlaw continuum longward of
1100\,\AA. They do not indicate any substantial reddening longward of
the break (except the 8 objects associated to class C, see BM05). In
Fig.\,\ref{fig:ext}, the extinction cross-section resulting from ISM
dust is represented by the silver continuous line.  The model depicted
consists of silicate and graphite grains with grain sizes ranging
between 50 and 2500\,\AA.


The extinction by SMC-like dust, however, is more peaky in the near-UV
and, as will be shown, cannot reproduce the sharp 1000\,\AA\ break
either. For comparison, the long dashed line in Fig.\,\ref{fig:ext}
represents the cross-section of a dust model consisting of relatively
small grain sizes, ranging from 3 to 300\,\AA, but with graphite
replaced by amorphous carbon (using results from Rouleau \& Martin
1991), that is, dust composed of 50\% silicate grains, 45\% amorphous
C grains and 5\% graphite grains. The dust-to-gas ratio is the same as
for the Galactic ISM model. The reason for replacing most of the `usual'
graphite grains by amorphous carbon is guided by the need to remove
the 2175\,\AA\ absorption feature present when small graphite grains
are abundant.  Our aim is to test an extinction curve that would be
SMC-like, in accordance with the extinction law inferred from the
SLOAN quasar sample by Richards et\,al. (2003), Hopkins et\,al. (2004)
and Willott (2005). In Fig\,\ref{fig:dust}, we indicate how the
previous powerlaw SED would look if absorbed by dust containing
silicates and amorphous carbon (corresponding to the long dashed line
extinction model in Fig.\,1), for the four columns values of
$N_{20}=0.5$, 1.0, 2.0 and 4.0, in units of $10^{20}$\,cm$^{-2}$.  A
sharp break does occur for columns $N_{20}\sim 2$. The break takes
place, however, at too long a wavelength in the near-UV. This is
illustrated in Fig\,\ref{fig:pg} where we compare the spectra of
PG\,1008+1319 with a powerlaw absorbed by nanodiamonds (gray
continuous line) and by silicate-amorphous carbon dust (gray dash
line). Clearly, only the nanodiamond dust provides a satisfactory
fit. 

Adopting the more commonly used SMC dust model of Pei (1992),
which consists only of silicate grains, would not improve the
situation, since the extinction cross-section is even shallower in the
near-UV in the Pei model than in the case of the silicate-amorphous
carbon dust. A comparison of these two extinction curves is made in
Fig.\,\ref{fig:ext}. The Pei cross-section has been mutiplied by $5.5$
Fig.\,\ref{fig:ext}) to show that it approximately overlaps in the
range 1800--5000\,\AA\ with the silicate-amorphous C extinction
curve. This confirms that the selective extinction of both dust models is
quite comparable in the optical domain.

\section{Conclusion}\label{sec:conclusion}

Dust absorption by nanodiamonds is successful in reproducing the 1000\,{\AA}
break as well as the far-UV rise seen at shorter wavelengths in distant
quasars. To confirm the possible existence of nanodiamond grains in
quasars, observations of the far-infrared emission bands caused by
hydrogenated nanodiamonds (van Kerckhoven et\,al. 2002; Jones et\,al. 2004) 
could be attempted or, alternatively, one may try to confirm a flux rise 
shortward of 700\,{\AA} for as many quasars as possible. The latter would
require high quality far-UV sensitive observations, which will hopefully be
provided by the World Space Observatory satellite (Barstow et\,al. 2003). 

The model presented is successful in both fitting the breaks shape and
the position of, as well as in reproducing the sharp flux recovery
observed in key quasar spectra, around 660\,{\AA}. While very high
redshift quasars ($z\ga 2.5$) appear not to be absorbed by dust, their
energy distribution suggests the existence of a higher energy
continuum rollover at $\simeq 670$\,{\AA}. As shown in BM05, when such
a break is incorporated into the description of the energy
distribution of all quasars, the dust absorption model can account for
the overall shape of the 'composite' spectral energy distribution of
the 184 quasars constructed by TZ02.

\acknowledgements The authors acknowledge support from CONACyT grant 40096-F.
HM acknowledges support by DFG grant Mu1164/5. 
We thank Randal Telfer for sharing the reduced HST-FOS spectra and
Diethild Starkmeth for proof reading.


\end{document}